\documentclass[twocolumn]{aastex63}
\usepackage{soul}

\newcommand\belo{\textit{B19}}

\received{November 11, 2019}
\accepted{February 21, 2020}
\submitjournal{ApJL}

\shorttitle{Splash without a merger}
\shortauthors{Amarante et al.}
\graphicspath{{./}{figures/}}
\begin{document}

\title{The Splash without a merger \footnote{Released on November, XXth, 2019}}

%% you desire. In v6.3 you can include a footnote in the title.

\correspondingauthor{Jo\~ao A. S. Amarante}
\email{joaoant@gmail.com}

\author{Jo\~ao A. S. Amarante}
\affiliation{Key Laboratory for Research in Galaxies and Cosmology\\ Shanghai  Astronomical Observatory, Chinese Academy of Sciences\\ 80 Nandan Road, Shanghai 200030, China}

\author{Leandro {Beraldo e Silva}}
\affiliation{Jeremiah Horrocks Institute, University of Central Lancashire, Preston, PR1 2HE, UK}

%\collaboration{1}{(AAS Journals Data Scientists collaboration)}

\author{Victor P. Debattista}
\affiliation{Jeremiah Horrocks Institute, University of Central Lancashire, Preston, PR1 2HE, UK}

\author{Martin C. Smith}
\affiliation{Key Laboratory for Research in Galaxies and Cosmology\\ Shanghai  Astronomical Observatory, Chinese Academy of Sciences\\ 80 Nandan Road, Shanghai 200030, China}



\begin{abstract}
The Milky Way's progenitor experienced several merger events which left their imprints on the stellar halo, including the {\it Gaia}-Sausage/Enceladus. Recently, it has been proposed that this event perturbed the proto-disk and gave rise to a metal rich ([Fe/H] $>-1$), low angular momentum ($v_{\phi} < 100$ km/s) stellar population. These stars have dynamical and chemical properties different from the accreted stellar halo, but are continuous with the canonical thick disk. In this letter, we use a hydrodynamical simulation of an isolated galaxy which develops clumps that produce a bimodal thin$+$thick disk chemistry to explore whether it forms such a population. We demonstrate clump scattering forms a metal-rich, low angular momentum population, without the need for a major merger. We show that, in the simulation, these stars have chemistry, kinematics and density distribution in good agreement with those in the Milky Way.%
\end{abstract}



%%%%%%%%%%%%%%%%% BODY OF PAPER %%%%%%%%%%%%%%%%%%

\section{Introduction}\label{sec:intro}
Selecting a pure sample of stars in either the  stellar halo or in the thick disk in the Solar Neighbourhood is complicated by their significant overlap in both their velocity and metallicity distributions. A common approach for selecting local stellar halo stars uses a kinematic cut to select high transverse velocity stars. By selecting stars with $v_t > 200$ km/s, one expects a negligible contamination by thick disk stars. However, with the high quality data provided by \textit{Gaia} \citep{gaia1, gaia2}, recent studies have shown that such a kinematic cut still leaves a population of stars with thick disk chemistry \citep{gaiahr, haywood, amarante}. This imprint is also seen when different kinematic criteria are chosen (e.g. \citealt{helmi2018}), challenging our understanding of the formation of the Milky Way's thick disk.\par
Furthermore, \citet{dimatteo2018} and \citet{amarante} found that many counter-rotating stars in the Solar Neighbourhood are too metal-rich to be considered part of the accreted halo. In particular, \citet{dimatteo2018} noted a low angular momentum population with thick disk chemistry (see, e.g., their figure 13) and proposed that the classical Milky Way (MW) inner-halo is actually composed of two stellar populations: \textit{i)} heated stars from the thick disk (referred by them as the ``Plume"), where the heating mechanism is associated with a major merger event named the \textit{Gaia}-Sausage/Enceladus\footnote{For short, we will just refer to it as the \textit{Gaia}-Sausage.}(\citealt{belo2018, helmi2018}); \textit{ii)} accreted stars from the \textit{Gaia}-Sausage.  \par
\citet{belo2019} (hereafter \belo), using the \citet{das} catalogue, disentangled the aforementioned low angular momentum structure from the canonical thick disk and stellar halo. They suggested that this structure, which they termed the Splash, is chemically and dynamically distinct from the known stellar populations in our Galaxy. Nonetheless, its formation must be linked to the thick disk as there is a smooth transition between the two populations in the kinematic-metallicity space. Moreover, they found Splash-like structures in hydrodynamical simulations where the host galaxy underwent a major merger. Therefore they concluded the proto-galactic disk of the MW was likely heated during the \textit{Gaia}-Sausage event about $10$ Gyr ago, in agreement with other studies \citep[e.g.][]{dimatteo2018, mackereth, gallart}, and thereby formed the Splash. Finally, they argued that the thick disk formation occurred before, during and up to $\sim$ 2 Gyr after the merger. \par 
On the other hand, \citet{clarke} presented a hydrodynamical simulation of an isolated galaxy that formed a thick disk purely via internal evolution driven by clump formation. These clumps dynamically heat the disk creating two chemically distinct disk components, with an overall double-exponential vertical profile \citep{beraldo}, similar to the MW's thin and thick disks. They also showed that the model's clumps are similar to those observed in high redshift galaxies.
Therefore a question that naturally arises is whether clumps are able to form stars with properties similar to the Splash, or whether the Splash stars uniquely need to form in a major merger event. In the following sections we demonstrate that clumps in the MW progenitor can produce stars with very similar kinematic and chemical properties as the observed Splash stars. This paper is organized as follows: Section \ref{sec:data} and \ref{sec:result} present the details of the simulated galaxy and shows its dynamical and chemical features, respectively. Section \ref{sec:conc} discusses the implications of our results.

\section{Simulation}\label{sec:data}
The simulation used in this paper is described in detail in \citet{clarke}. The initial conditions are characterized by a spherical hot gas corona embedded in a dark matter halo with a Navarro-Frenk-White \citep{NFW} density profile with virial radius $r_{200}\approx 200$ kpc and mass of $10^{12}\text{M}_\odot$. The gas has an initial net rotation and cools via metal-line cooling \citep{shen2010}. It settles into a disk, and stars form wherever the temperature drops below 15,000 K and the density exceeds 1 $cm^{-3}$. Feedback by supernova explosions follows the blastwave implementation of \citet{stinson}, with thermal energy being injected to the interstellar medium with an efficiency of $10\%$. The feedback of asymptotic giant branch stars is also taken into account. Gas phase diffusion uses the method of \citet{shen2010}.

These initial conditions are self-consistently evolved for 10 Gyrs with the smooth particle hydrodynamics+N-body tree-code {\sc gasoline} \citep{wadsley}. As shown by \cite{clarke}, the metal-line cooling results in the formation of clumps during early times of the simulated galaxy. At the end of the simulation, the galaxy presents a chemical bi-modality and geometric properties very similar to those observed for the MW \citep[as shown by][]{beraldo}. In particular, a thick disk, composed of old, $\alpha$-rich stars is formed\footnote{In the simulation, the [$\alpha$/Fe] abundance is tracked by [O/Fe].}. This simulation evolves as an isolated galaxy, i.e. without any merger that could produce the metal-poor and high radial velocity dispersion population associated with the \textit{Gaia}-Sausage observed for the MW \citep[see][]{belo2018, helmi2018} or any stellar halo component.

For comparison, we also ran another simulation from the same initial conditions and with the same procedure, except for the supernova feedback efficiency, which was set to $80\%$. This high feedback efficiency inhibits the formation of clumps and the simulation does not form a chemical or a geometric thick disk \citep{beraldo}. We refer to this simulation as the non-clumpy simulation. This allows us to study the role of the clumps on the formation of the different populations in the simulation.
\subsection{Simulated Solar Neighbourhood}\label{sec:SN}
In order to reduce contamination from thin disk stars, \belo\,selected stars in the region $0.5< |z|/\mathrm{kpc}<3$. For reliable comparisons to their results, we apply a similar cut to the simulation data and define our mock Solar Neighbourhood as all star-particles in the region $5 < R/\mathrm{kpc} < 11$ and $|z| < 3$ kpc, where $R$ is the cylindrical radius centered at the simulated galaxy center, and $z$ is the height from the simulated galaxy plane. Unless explicitly mentioned, all our results are based on this geometrical slice. We have tested using the same $|z|$ range as \belo\, and the difference is the reduced amount of thin disk star-particles. Nonetheless the trends between the simulated thick disk and Splash region discussed below are the same. Finally, throughout the paper the velocities are given in cylindrical coordinates centered on the simulated galaxy center; $v_{R}$, $v_{\phi}$ and $v_{z}$ are positive in the direction of the galaxy centre, galaxy rotation and angular momentum vector, respectively. 

\section{Results}\label{sec:result}

\begin{figure*}
    \includegraphics[scale=0.43]{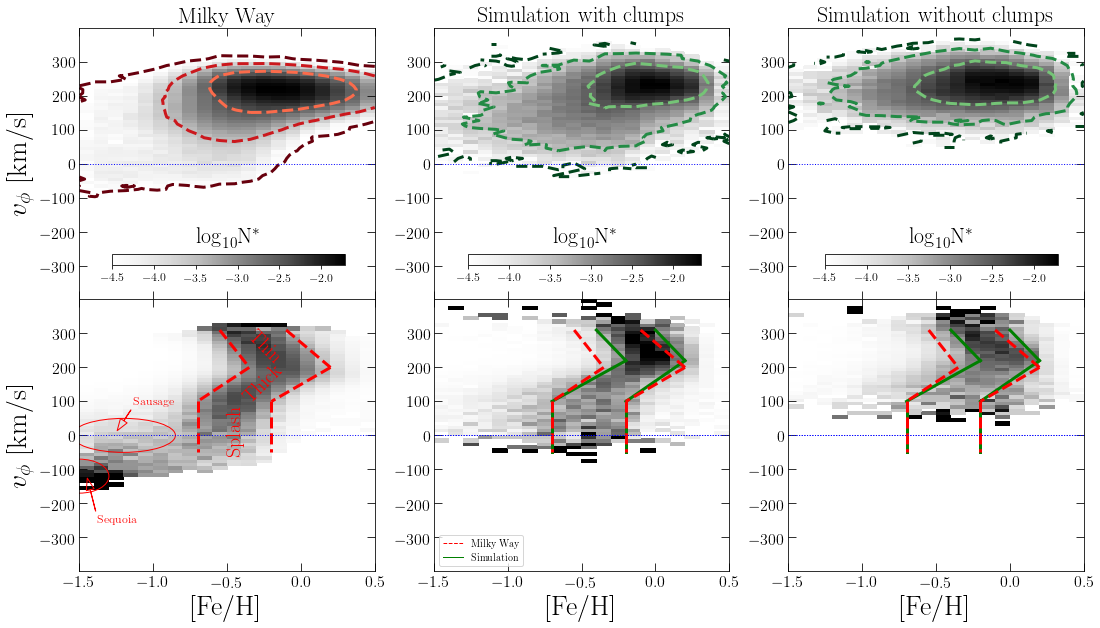}
    \caption{Left column shows the MW stars with $0.5< |z|/\mathrm{kpc}<3$ from \citet{das}. The simulation panels are for stars with $5< R/\mathrm{kpc} < 11$ and $|z| < 3$ kpc, see details in the text. The middle and right column shows data from the clumpy and non-clumpy simulation, respectively. \textbf{Top row:} absolute density plots where each bin is divided by the total number of stars/star-particles following the selection criteria. The dashed-lines are density contours on a logarithmic scale. \textbf{Bottom row:} Row-normalised 2D density plot in the [Fe/H]-$v_{\phi}$ plane. The dashed red lines indicate the location of the thin disk, thick disk and Splash, as defined by \belo\,in the observed data. The solid green lines show the equivalent regions in the clumpy simulation. The Splash region produced by the simulation is similar to that of the MW. Comparison of the central and right panels shows that the clumps are necessary to produce the Splash stars. Moreover the simulation without clumps shows no Splash region in the velocity-metallicity plane. Finally, we also illustrate in the left panel the two known accreted stellar halo structures: \textit{Gaia}-Sausage and Sequoia.}
    \label{fig:row_normal}
    \end{figure*}
We now compare the kinematics and chemistry of the model to the results of \belo. As described in Section \ref{sec:data}, our simulation self-consistently evolves as an isolated galaxy and thus has no accreted stellar halo. Therefore it has a lower fraction of metal-poor and of counter-rotating stars than the MW. For better comparison with the observational results, in most of our figures we adopt similar scales and colour schemes as those used by \belo. Moreover, whenever we use data from \citet{das}\footnote{\url{http://www.ast.cam.ac.uk/~jls/data/gaia_spectro.hdf5}} we select stars with the same spatial selection, small velocity errors ($\sigma_{v_{\phi}} < 20$ km/s), low [Fe/H] uncertainty ($\sigma_{\rm [Fe/H]} <0.15$) and accurate parallax ($\varpi/{\sigma_{\varpi}} > 5$) as in \belo.\par
\subsection{Chemistry-kinematics features}
Fig. \ref{fig:row_normal} top row shows the absolute density plot in the [Fe/H]-$v_{\phi}$ plane for both the MW (left) and the simulations (center and right). Each bin is divided by the total number of stars/star-particles following the selection criteria in the MW/simulations. As already discussed in \belo, the MW has a significant amount of metal-rich ([Fe/H]$>-1$), low angular momentum ($v_{\phi} <100$ km/s) stars (see, e.g., \citealt{nissen, fernandez2019}). The simulation with clumps also has a significant amount of such metal-rich low angular momentum stars, but lacks a significant amount of retrograde stars, mainly due to the absence of accreted stars\footnote{We note that not all retrograde stars with [Fe/H]$>$-1 observed in the MW are accreted.}, and has a slight overdensity at [Fe/H]$>0$ and $v_{\phi} <100$ km/s compared to MW (but see \belo\, figure 1, where, e.g., the left panels clearly shows stars in this region). For comparison, the simulation without clumps, right panel, lacks a significant low angular momentum population.  \par
The bottom row of Fig. \ref{fig:row_normal} shows the row-normalised density plot in the [Fe/H]-$v_{\phi}$ plane for both the MW (left) and simulations (center and right). This normalisation has the advantage of enhancing the known velocity-metallicity correlation for both the thin and thick disk, where the former (\textit{latter}) has a negative (\textit{positive}) $v_{\phi}$ gradient with [Fe/H]. The red contours in the left panel indicate the different populations in the MW, as identified by \belo, where the Splash is defined as the over-density seen at the low angular momentum ($v_{\phi} \leq 80$ km/s) and relatively metal-rich ($-0.7<$[Fe/H]$<-0.2$) region. The two known stellar halo over-densities, \textit{Gaia}-Sausage \citep{belo2018, helmi2018} and Sequoia \citep{barba, myeong}, are also indicated. Although the Splash region is defined ad hoc in this plot, it is significantly different from the accreted stellar halo and the classical thick disk (defined as the $\alpha$-rich disk, e.g. \citealt{bensby,hayden}) regions, as shown in \belo\,and later in this work.\par
The bottom-middle panel of Fig. \ref{fig:row_normal} shows the results from the simulated clumpy galaxy. The same trends observed in the MW's thin and thick disk are also observed here. In this panel, the red dashed lines are those defined  by \belo\, with the MW data, while the green solid lines are defined from the simulation. The narrowness of the bands for the thin and thick disks in the simulation can be understood as being due to the absence of any observational errors. Nonetheless, the Splash region is exactly the same both in the MW and in the simulation. In the right panel, we show the same plot for the non-clumpy simulation. The solid and dashed red lines are the same as in the middle panel. The trends for the thin and thick disk are not as clear now and this galaxy can be described as having only one disk component \citep{clarke, beraldo}. Moreover, it lacks the low angular momentum ($v_\phi \lesssim 50$ km/s) star particles of the Splash region. \par
\begin{figure*}
    
    \includegraphics[scale=0.43]{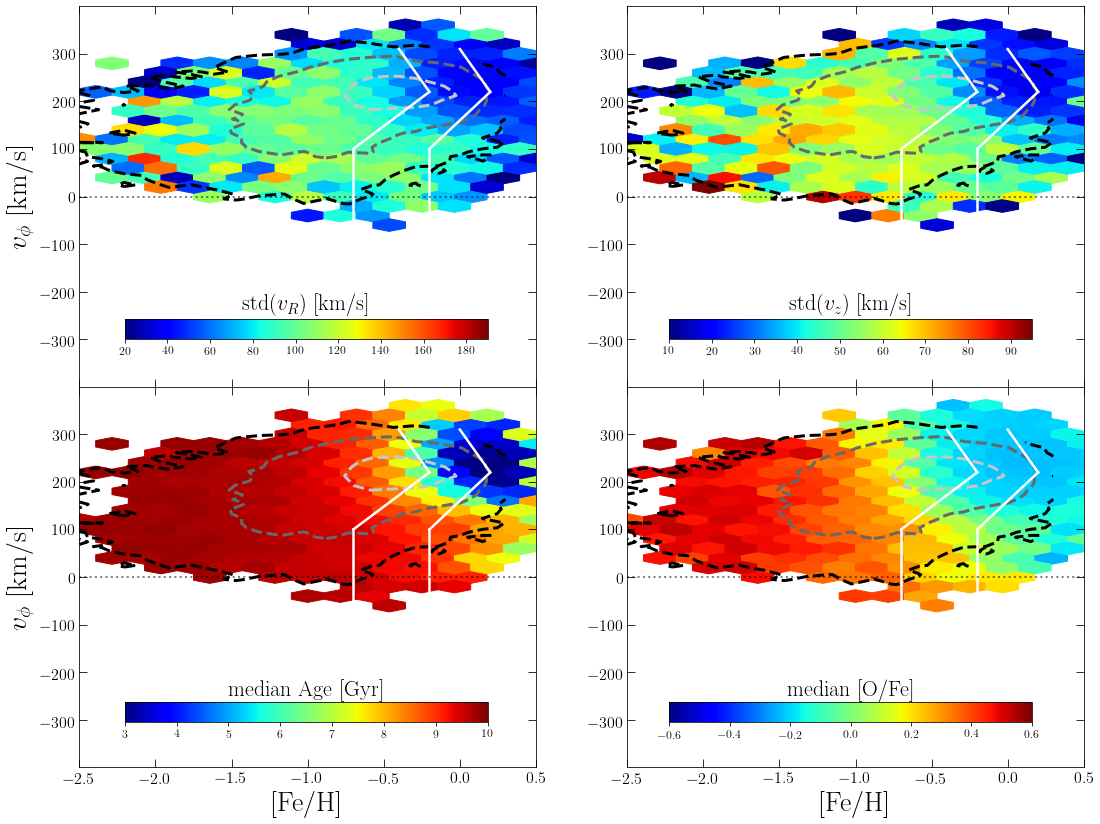}
    \caption{[Fe-H]-$v_{\phi}$ plane in the simulated clumpy galaxy, $5< R/\mathrm{kpc} < 11$ and $|z| < 3$ kpc, which was evolved for only 10 Gyr. The dashed lines are the density contours on a logarithmic scale. The solid white lines define the thin disk, thick disk and Splash regions in the simulation as in Fig. \ref{fig:row_normal}. \textbf{Top row:} on the left, colour-coded by radial velocity dispersion, it is possible to see the transition from the thin disk population towards the thick disk going from mild to higher radial velocity dispersion. The Splash region has a higher radial velocity dispersion compared with the thick disk counterpart. A similar trend is observed in the vertical velocity dispersion shown on the right. \textbf{Bottom row:} on the left we see that the Splash region is composed of the oldest stars in the simulation, whereas the thin disk is the youngest and the thick disk has intermediate to old stars. Finally on the right, the smooth [O/Fe] transition from the thick region to the Splash region hints at their related origins.
    }
    \label{fig:leandro_sim}
\end{figure*}
Fig. \ref{fig:leandro_sim} shows the [Fe/H]-$v_{\phi}$ plane for the simulated MW analogue colour-coded according to different properties (see figure 1 of \belo\,for comparison). The top left panel is colour-coded by dispersion in $v_R$ for each bin. As can be noted, the simulated galaxy has the same trends as in the MW: the thin disk has a mild $v_R$ dispersion and it smoothly increases from the thick disk towards the Splash region. As expected, the large $v_R$ dispersion stars observed for [Fe/H] $<-0.7$, associated with the \textit{Gaia}-Sausage and the Sequoia in the MW, are not present in our simulation as these stars were accreted. As in the MW, a similar trend is also observed for the dispersion in $v_z$, see top right panel. We stress that not only do trends match between the simulated galaxy and the observed data, but also the velocity dispersion values are very similar.\par
The bottom panels in Fig. \ref{fig:leandro_sim} show how the median age and [O/Fe] vary in the [Fe/H]-$v_{\phi}$ plane. As in the MW, the thin disk region is mostly comprised of young stars and the disk gets older in its thick component. The oldest stars are located in the Splash region. The thin disk region in our simulation is younger compared with the results in \belo; as mentioned in Section \ref{sec:data}, the simulation is evolved for only 10 Gyr and thus it will not reflect the same oldest ages as observed in the MW. Nonetheless, the thick disk and Splash region still have similar age gradients compared to the observations. Finally, the trends in [O/Fe] are nicely matched to the [$\alpha$/Fe] trends in the MW, where there is a smooth transition from slightly [O/Fe]-rich for the thick disk to higher values in the Splash region. \par
\begin{figure*}
    
    \includegraphics[width=16cm]{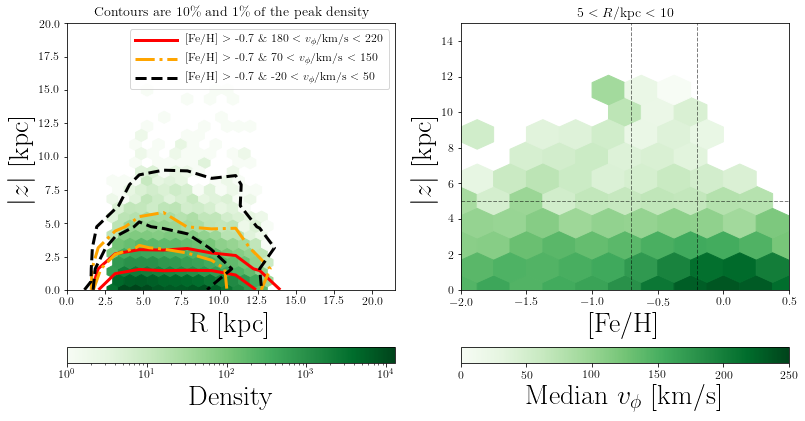}
    
    \caption{The spatial extent of the Splash stars in the simulation. The left plot shows the distribution of the metal-rich stars for three different velocity ranges. The thin and thick disk-like stars are indicated by the red and orange curves, respectively. The curves are the contours of $10\%$ and $1\%$ of the peak density. The lower angular momentum stars (dashed black curves) represent the Splash stars and have similar $|z|$ extension in comparison to observations (see \belo\,figure 8). On the right, we show the [Fe/H]-$|z|$ plane colour coded by median $v_{\phi}$. The trends are also remarkably similar to the observations; the transition between the canonical thick-disk to the Splash-dominated region also occurs at $|z| \sim 5$ kpc.}
    \label{fig:z_exten}
\end{figure*}
\citet{clarke} showed that, in the simulation, a small population of $\alpha$-poor stars forms at the same time as the $\alpha$-rich ones (indeed this is one of the key predictions of the model). We have verified that the
publicly available APOGEE-Gaia DR12  catalogue\footnote{\url{https://www.sdss.org/dr12/irspec/spectro_data/}} indeed confirms that retrograde Splash stars are found both in the $\alpha$-rich and $\alpha$-poor in-situ regions (as defined in \citealt{mackereth}), supporting our conclusion that scattering by clumps can produce the counter-rotating Splash stars in the MW. \par 
\belo\, analysed LAMOST K giants \citep{Luo2015} in order to verify how far from the plane the Splash stars are detected and to determine where the canonical thick disk transitions to the Splash-dominated region. They found that at $|z| \sim 5$ kpc there is a sharp gradient in median $v_{\phi}$ where the low angular momentum Splash stars dominate for higher $|z|$ (see their figure 8). Although the stellar density of the disk in our simulated galaxy decreases faster with $|z|$ compared with the MW's disk, we are still able to verify the transition region, if any, in the simulation. The left panel in Fig. \ref{fig:z_exten} shows the stellar density in the $R-|z|$ plane for the simulation. The red and orange curves have the same velocity-metallicity intervals as in \belo\, and they are a rough representation of the canonical thin and thick disks, respectively. The black dashed curves outline the spatial extent of the Splash stars in our simulation. In order to avoid the tail of the stellar halo distribution, \belo\,restricted their velocity range to $ -20< v_{\phi} < 50$ km/s. Even though our simulation does not have an accreted halo, we adopt the same range. We note that both in the simulation and in the MW, the Splash stars extend up to $|z| \sim 10$ kpc and are also concentrated within $R \sim 10$ kpc.   \par
The right panel in Fig. \ref{fig:z_exten} shows how the median $v_{\phi}$ changes in the [Fe/H]-$|z|$ plane. In the Splash metallicity range, defined by the two vertical dashed lines, we note two features similar to those observed in the MW: \textit{i.} the negative gradient of thick disk's median $v_{\phi}$ with $|z|$; \textit{ii.} a transition at $|z| \sim 5$ kpc (horizontal dashed line) where the low angular momentum Splash stars start to dominate for larger heights. This reinforces the idea that internal dynamical processes in a clumpy proto-disk can also create Splash stars. \par
\subsection{Simulated Splash number fraction}\label{sec:fraction}
\begin{figure*}
    \includegraphics[scale=0.43]{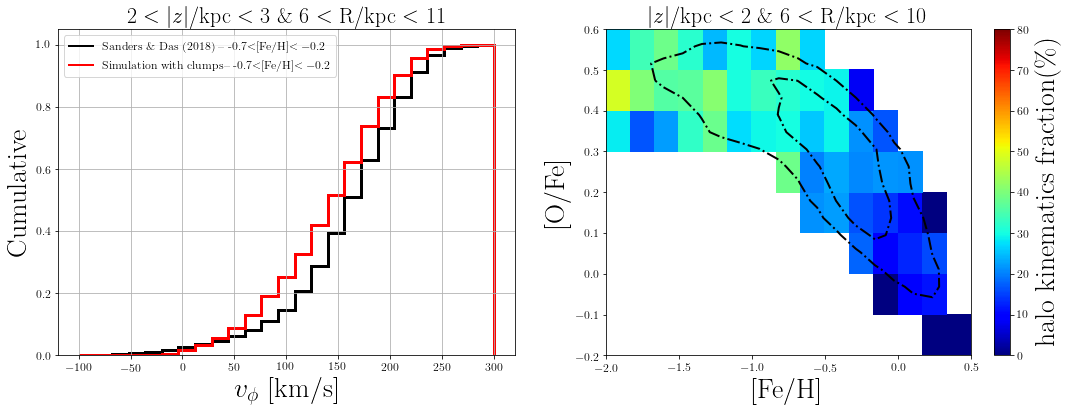}    
  \caption{\textbf{Left panel:} The cumulative $v_{\phi}$ distribution for the MW (black) and simulation with clumps (red) in the Splash metallicity regime, $-0.7<$[Fe/H]$<-0.2$. The fractions of stars and star-particles in the tail of the distribution are $\approx 15\%$ and $\approx 25\%$, respectively. \textbf{Right panel:} An estimate of the fraction of star-particles with halo like kinematics, $\sqrt{(v_{\phi}-v_{LSR})^2 + v^2_R + v^2_z} > 180$ km/s, following the criteria from \citet{dimatteo2018} (see their figure 18 for a comparison with the MW). We show the fractions for the chemically defined thick disk in the simulation \citep{clarke}. The black curves are the contours of $50\%$ and $10\%$ of the peak density. The trends in the simulation with clumps are very similar to the MW, see text for details. }
    \label{fig:fraction}
    \end{figure*}

In this subsection, we estimate the Splash-like population in the simulation with clumps and compare with the MW. \belo\, estimated that in the range $2 <|z|$/kpc$<3$ and $-0.7<$[Fe/H]$<-0.2$ the Splash population dominates the tail of the $v_{\phi}$ distribution, i.e., $v_{\phi}<100$ km/s. The left panel of Fig. \ref{fig:fraction} shows the cumulative distribution of $v_{\phi}$ for the MW (black) and simulation (red) in the aforementioned metallicity and spatial range: approximately $15\%$ of the stars in the MW have $v_{\phi} < 100$km/s, whereas the simulation with clumps has about $25\%$ of star-particles in the same regime. However, if we scale $v_{\phi}$ by the median value for each distribution, the simulation with clumps has $\approx 20\%$ of star-particles in the equivalent low angular momentum tail, remarkably similar to the MW. \par
The right panel of Fig. \ref{fig:fraction} shows the fraction of star-particles with halo kinematics in the simulation chemical thick disk. We used a spatial range similar to that used in \citet{dimatteo2018}. We also followed their criterion for halo kinematics, $\sqrt{(v_{\phi}-v_{LSR})^2 + v^2_R + v^2_z} > 180$ km/s, where $v_{LSR}=233$ km/s and it is the velocity at the local standard of rest in the simulation \citep{clarke}. The separation criterion between the thin and thick disks for the simulations follows \citet{clarke}. We observe that the simulation follows the same trends as in the MW (see figure 18 in \citealt{dimatteo2018}): \textit{i.} the fraction of halo-like kinematics is $\approx 25 \%$ in the range where the Splash stars are concentrated; \textit{ii.} this fraction drops quickly as the metallicity increases. The larger fraction of halo-kinematics stars at [Fe/H]$>$-0.2 in the simulation, compared with the MW, can be due to differences in the details of the later star formation history and/or the selection function of the  APOGEE data.

\subsection{The age of simulated Splash stars}\label{sec:age}
In order to compare the age difference between the accreted halo and the Splash stars, \belo\,analysed the age distribution of the counter-rotating stars for two distinct metallicity ranges (see their figure 4). The metal-poor ([Fe/H]$<$-0.7) region, corresponding to the accreted halo has a peak at $\sim 12.5$\footnote{Their ages are limited to 12.5 Gyr.} Gyr and a sharp drop at $\sim 10$ Gyr. On the other hand, the age distribution for the Splash stars, -0.7$<$[Fe/H]$<$-0.2, has a slightly younger peak at $\sim 11.5$ Gyr and also a sharp drop at $\sim 10$ Gyr ($\sim 30\%$ of these stars are younger than 10 Gyr). They propose that the Splash is made up of stars from the Galaxy's proto-disk whose orbits were heated by to the \textit{Gaia}-Sausage accretion event. Moreover, the star formation for the Splash stars ceased just after the merger, i.e. around 9.5 Gyr ago. \par
\begin{figure*}
    \includegraphics[scale=0.43]{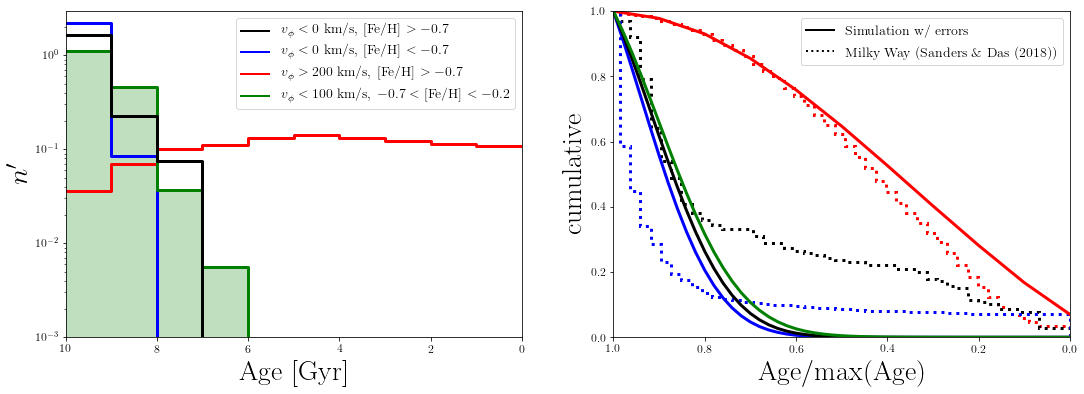}
    \caption{\textbf{Left panel:} Error free age distribution for the simulated stars in the simulated clumpy galaxy with $5< R/\mathrm{kpc} < 11$ and $|z| < 3$ kpc. $n'$ is the probability density function of each 1 Gyr age bin. The black and blue curves show the counter-rotating stars for metal-rich and metal-poor intervals, respectively. All the counter-rotating stars in the simulation are older than 7 Gyr. The absence of the shift in the peak of the age distribution between the two curves, seen by \belo, reflects the fact that our simulation has no accreted halo. The red curve shows the age distribution of the classical thin$+$thick disk in the simulation, which was only evolved for 10 Gyr. The green shaded area corresponds to the age distribution of the simulation Splash region defined in Fig. \ref{fig:row_normal}. As it includes stars with $v_{\phi}<$100 km/s, some thick disk stars are present. \textbf{Right panel}: cumulative age distribution for the MW (dotted lines) and simulation (solid lines) convolved with observational errors. The colours correspond to the same velocity and [Fe/H] interval as in the left panel. We divide each age by the maximum age in the MW data (12.5 Gyr) and the simulation (10 Gyr). }
    \label{fig:age_hist}
\end{figure*}

In the left panel of Fig. \ref{fig:age_hist}, we show the age distribution in our simulation using the same velocity-metallicity interval as in \belo. As already mentioned, our oldest age is 10 Gyr and the fact there is no shift between the metal-poor (blue) and Splash stars (black) curves merely reflects that our simulation represents an isolated galaxy with no old accreted stellar halo. Nonetheless, it is evident that the counter-rotating stars in the simulation are also old and cease to form after 3 Gyr. This fact is qualitatively similar to what is observed in the MW. The red curve in this panel corresponds to the canonical thin$+$thick distribution, $v_{\phi}>200$ km/s and [Fe/H]$>$-0.7. As expected it extends over the full age range in the simulated galaxy in agreement with the observed data. Finally, the green shaded area is the age distribution for the Splash stars defined as in Fig. \ref{fig:row_normal}, i.e. $v_{\phi}<100$ km/s and $-0.7<$[Fe/H]$<-0.2$. Now, the age extends to an extra 1 Gyr, due to the contamination of thick disk, i.e. $\alpha$-rich stars. As shown by \citet{clarke}, all the star-particles in the simulation thick disk are older than 6 Gyr. Therefore, our result reinforces the connection between the thick disk and the Splash stars, where the latter could naturally exist in galaxies with a disk dichotomy. \par
Fig. \ref{fig:age_hist} right panel shows the cumulative distribution for the MW (dotted lines) and the simulation (solid lines) convolved with the median error from the observational data (median $\sigma_{age} \approx 1.3$ Gyr)\footnote{We select stars in \citet{das} with estimated age error lower than 2 Gyrs.}. For a better comparison, we divided the age intervals by the maximum age in the simulation (10 Gyr) and observation (12.5 Gyr).
For $v_{\phi} < 0$ and [Fe/H]$>-0.7$ the MW (black dotted line) has a truncated star formation history characterized by: \textit{i)} approximately $60\%$ of stars are older than $age/max(age) \approx 0.8$, \textit{ii)} an extended tail towards young ages. \belo\, associated the truncation with the cessation of the MW's disc heating during the \textit{Gaia}-Sausage merger. Similarly to the MW, the simulation also has approximately $60\%$ of stars older than $age/max(age) \approx 0.8$ (black and green solid lines). The lack of young Splash-like stars in the simulation is explained by the fact that the clumps, which are responsible for producing the Splash, stop forming after 4 Gyr. The difference between the simulation and the MW, in terms of the extended tail of ages, only happens after both the clumps and the {\it Gaia}-Sausage are completed. 
We also note, by analysing the simulation snapshots, that this old population is already kinematically hot in the first 3 Gyrs of the simulation. Therefore, internal dynamical heating in a MW-like galaxy is enough to heat the primordial disk in its early stages and produce a kinematically hot population with $v_{\phi} < 100$ km/s, as shown by the green distribution in Fig. \ref{fig:age_hist}.    \par

\section{Conclusions}\label{sec:conc}
\citet{dimatteo2018} suggested that the Splash stars (which they refer to as the ``Plume") represent the low angular momentum tail of the thick disk and therefore it is not a distinct component from the MW. On the other hand, \belo\, argues on the necessity of three components (thin disk, thick disk and Splash) to explain the $v_{\phi}$ distribution for $-0.7<$[Fe/H]$<-0.2$ and $2 < |z|/kpc <3$. Based on their age distribution, they also suggest that the Splash formed only in the early times of the MW, whereas the thick disk had a continuous formation over time.\par
From \belo's work the existence of four distinct stellar populations in the dynamical-chemical space is suggested: thin disk, thick disk, accreted halo and the Splash, the latter sometimes being referred to as the in-situ halo. The origin of the Splash also relates to the thick disk and it possibly contains the first stars in the proto-disk of our Galaxy's progenitor. Besides that, there is clear evidence of a major merger event that occurred in the early stages of the MW (e.g. \citealt{amina, belo2018, haywood, dimatteo2018}). \citet{dimatteo2018} and \belo\, proposed that this merger event could have excited the proto-disk and given rise to the low angular momentum, high radial velocity dispersion and relatively metal-rich Splash stars. \par
In this letter we analysed an isolated galaxy simulation in order to investigate the formation of Splash stars. We showed that a clumpy MW analogue can naturally form not just the chemical and geometric thin and thick disks \citep{clarke, beraldo}, but also the Splash, with distribution, kinematics and chemistry similar to those observed in our Galaxy. This is a new scenario for the formation of the Splash, as in our case there was no accretion event and therefore no accreted stellar halo. Moreover, in this scenario, the formation of the Splash stars occurs in the simulation's first Gyrs. The thick-disk and Splash population have a common origin, where the latter is the low angular momentum tail of the former, as initially suggested by \citet{dimatteo2018}. This is due to the smooth transition in kinematics, age and [O/Fe] between the thick-disk and Splash region seen in Fig. \ref{fig:leandro_sim}. We also show that a similar simulation without clumps fails to reproduce the low rotational velocity patterns of the MW. We verified that the simulated Splash has a similar number fraction (Section \ref{sec:fraction}) and age trends (Section \ref{sec:age}) as observed in the MW. Our results suggest that a Splash population is expected in any galaxy which underwent a clumpy star-formation episode.\par
Finally, we note that the two different formation scenarios for the Splash population, i.e., clumpy star formation (this work) or a major merger (\citealt{dimatteo2018, belo2019}), are not mutually exclusive, since clumps can have an ex-situ origin associated to mergers (e.g. \citealt{mandelker}). The effect of the merger on the disk depends on several parameters, such as the initial gas fraction and orbit (see e.g. \citealt{dimatteo2011}). Moreover, the proximity of the satellite's disruption, whether close or far from the disk, also plays an important role on the disk heating \citep{sellwood}, and requires further study for the {\it Gaia}-Sausage.
\acknowledgments
The authors wish to thank the anonymous referee for useful comments that helped improve this work. J.A. and M.C.S acknowledge support from the National Key Basic Research and Development Program of China (No. 2018YFA0404501) and NSFC grant 11673083. J.A. also acknowledges The World Academy of Sciences and the Chinese Academy of Sciences for the CAS-TWAS fellowship. J.A. is also grateful for the hospitality of Jeremiah Horrocks Institute at UCLan during his visit. V.P.D. and L.B.S. are supported by STFC Consolidated grant \#ST/R000786/1. The simulations in this paper were run at the DiRAC Shared Memory Processing system at the University of Cambridge, operated by the COSMOS Project at the Department of Applied Mathematics and Theoretical Physics on behalf of the STFC DiRAC HPC Facility (\url{www.dirac.ac.uk}). This equipment was funded by BIS National E-infrastructure capital grant ST/J005673/1, STFC capital grant ST/H008586/1 and STFC DiRAC Operations grant ST/K00333X/1. DiRAC is part of the National E-Infrastructure.

%\vspace{5mm}
%\facilities{HST(STIS), Swift(XRT and UVOT), AAVSO, CTIO:1.3m,


%\software{astropy \citep{2013A&A...558A..33A},  
       %   Cloudy \citep{2013RMxAA..49..137F}, 
        %  SExtractor \citep{1996A&AS..117..393B}
         % }


%\appendix

%\bibliography{sample63}{}
\bibliography{ref}{}
\bibliographystyle{aasjournal}



\end{document}